\documentstyle[aps,prl,multicol,epsf]{revtex}
\begin{document}
\draft
\title
{Noiseless Quantum Codes}
\author {P. Zanardi $^{1,2}$ and M. Rasetti $^{2,3}$}
\address{$^{1}$ ISI Foundation, Villa Gualino, Torino \\
 $^2$ Unit\`a INFM, Politecnico di Torino,\\
$^3$ Dipartimento di Fisica, Politecnico di Torino\\
Corso Duca degli Abruzzi 24, I-10129 Torino, Italy
}
\maketitle
 \begin{abstract}
{
In this paper we study a model  quantum register $\cal R$ made of
$N$ replicas (cells) of a given finite-dimensional quantum system S.
Assuming that all  cells are coupled with a common  environment with equal strength
we show that, for $N$ large enough, in the Hilbert space of $\cal R$ there exists
a linear subspace ${\cal C}_N$ which is dynamically decoupled from the environment.
The states in ${\cal C}_N$  evolve unitarily  and are therefore 
decoherence-dissipation  free.
The space ${\cal C}_N$ realizes a noiseless quantum code in which
 information can be stored, in  principle, for arbitrarily 
long time without being affected by errors.  
}
\end{abstract}
\begin{multicols}{2}[]
\narrowtext
Since the early days of quantum computation \cite{QC} theory
it has been clear that maintaining quantum coherence
in any computing system is an essential requirement
in order to fully exploit the new possibilities opened by quantum mechanics.  
This issue is known as the {\sl decoherence} problem \cite{ZUDI}.
Indeed, any real-life device unavoidably interacts with its environment,
wich is, typically, made by a huge amount of uncontrollable degrees of freedom.
This interaction causes a  corruption of the information stored in the system
as well as errors in computation steps, that  eventually lead
to wrong outputs.  
One of the possible approaches to overcome such difficulty,
in  analogy with classical computation, is to resort to redundancy
in encoding information, by means of the so-called 
{\sl error correcting codes} (ECC). 
In these schemes -- pionereed in \cite{ERROR} and raised
to a high level of mathematical sophistication in \cite{ERROR1} --
   information is encoded in  linear subspaces $\cal C$ 
(codes)
of the total Hilbert space in such a way that 'errors' induced
by the interaction with the enviroment can be detected and corrected.
The essential point is that the detection of errors, if
they belong to  the class of errors
correctable by the given code, should be performed without gaining any information
about the actual state of the computing system prior to corruption.
Otherwise this would result in a further decoherence.
The ECC approach can thus be considered as a sort of {\sl active} stabilization
of a quantum state, in which  by monitoring the system
and conditionally carrying on suitable operations, one
 prevents loss of information. 
The typical system considered in quantum-information context
is a $N$-{\sl qubit register} $\cal R$ made of $N$ replicas of a two-level
system S (the qubit). 
In the ECC literature, once more in analogy with the classical case,
it is assumed that  each qubit of $\cal R$ is coupled
with an independent environment.
\\In this letter we will show that the so far neglected case
in which all the qubits can be considered symmetrically coupled
withe same environment might  provide a new strategy
in the struggle for  preserving quantum coherence.   
The idea  is that, in the presence of a such 'coherent' environmental noise,
 one  can  design states  that are hardly
corrupted rather than states that can be easily corrected. 
In other words, the present approach consists in a {\sl passive} (i.e. intrinsic) 
stabilization of quantum information,
and in this sense it is complementary to EC. The resulting
codes could be called {\sl Error Avoiding}.
Furthermore, from the broader point of view of the  theory of open
 quantum systems,
our result shows a systematic way of building  non-trivial models 
in which dynamical symmetry  allows unitary evolution of a subspace
while the remaining part of the Hilbert space gets strongly entangled with the environment.
In the following we first briefly recall the basic mechanism of  decoherence.
If ${\cal H}_S,\,{\cal H}_B$ denote, respectively, the system and the environment 
 Hilbert
spaces, the total Hilbert space is given by the tensor product ${\cal H}={\cal H}_S\otimes{\cal H}_B.$
Let $\rho_S$ ($\rho_B$) be a state over ${\cal H}_S$ (${\cal H}_B$) (i.e.
$\rho_\alpha
\in\mbox{End}({\cal H}_\alpha),\,\rho_\alpha=\rho_\alpha^\dagger,\, \rho_\alpha\ge 0,\,\mbox{tr}
(\rho_\alpha)=1,\,
\alpha=S,\,B$).
According to quantum mechanics,  time evolution of the overall (closed) system is unitary, therefore
if $\rho(0)=\rho_S\otimes\rho_B$ is the initial state, then for any $t\ge 0$ one has
$\rho(t)=U_t\,\rho(0)\,U^\dagger_t,$
($U_t^{-1}=U_t^\dagger$). The induced (Liouvillian) evolution on ${\cal H}_S$ (open) is given by
$L^{\rho_B}_t\colon\rho_S\rightarrow\mbox{tr}^B\,\rho(t),$ 
where $\mbox{tr}^B$ denotes  partial trace over ${\cal H}_B.$
The crucial point is that
even if $\rho_S$ is a pure state ($\rho_S^2=\rho_S$), in a very short time
it gets  entangled with the bath
and becomes mixed ($\rho_S^2\neq\rho_S$). 
Typically,
in a suitable ${\cal H}_S$-basis, the off-diagonal elements of $\rho_S$
behave like $\exp(-t/{\tau_{Deco}}).$ The energy $\hbar \tau_{Deco}^{-1}$ is a measure of the rate 
at wich the information loss occurs. If an EC strategy is not used
$\tau_{Deco}$
 sets an upper bound to the duration of any reliable 
computation.   Notice that this mechanism, due to quantum fluctuations,
 is active at finite as well as at
zero temperature and does not necessarly imply that dissipation takes place.
Let us then 
begin by considering a simple example, important for quantum information applications --
$N$ identical  two-level systems
($N$-qubit register)
coupled with a {\sl single}  thermal bath  described by a collection of non-interacting linear oscillators.
The Hamiltonian of the register (bath) is given by $H_S=\epsilon \sum_{i=1}^N\sigma^z_i,$
($H_B=\sum_k\omega_k\,b_k^\dagger\,b_k$).
The bath-register interaction Hamiltonian is
\begin{equation}
H_I=\sum_{k,i=1}^N (
g_{ki}\, \sigma^+_i\,b_k +f_{ki}\,  \sigma^+_i\,b^\dagger +h_{ki}\, \sigma_i^z b_k +\mbox{h.c.}).
\label{IntSpin}
\end{equation}
The operators $\{\sigma_i^\alpha\}$ span $N$ local $sl(2)$ algebras
\begin{equation}
[\sigma_i^z,\,\sigma^\pm_j]=\pm\delta_{ij}\sigma^\pm_i,\; [\sigma^+_i,\,\sigma_j^-]=2\,\delta_{ij} \sigma_i^z.
\label{sl2}
\end{equation}
These commutation relations make clear the physical meaning of the interaction (\ref{IntSpin}) in terms
of elementary processes:
the first (second) term describes the excitation of the qubit by the absorbtion (emission)
of a bath mode with probability amplitude $f_{ki}$ ($g_{ki}$). This (togheter with the conjugate processes)
is the dissipative part of the interaction,  responsible for the
 (irreversible) exchange
of energy between  register and  bath.
The third term in equation (\ref{IntSpin}) is a conservative coupling
that induces pure dephasing between states corresponding to different eigenvalues 
of operators $\{\sigma_i^z\}.$
Now we make the basic physical assumption:
the coupling functions $g_{kj}, f_{kj}, h_{kj}$ {\sl do not
depend on the
replica index $j.$}
This is a generalization of the Dicke limit of quantum optics \cite{HELI}.
Such an  assumption can be justified
 if the replicas have very close  spatial positions
with respect to  the bath  coherence length $\xi_C$.
Indeed if, for istance,  $g_{kj}=g_k e^{i\,k\,R_j}$
($\{R_j\}$ denoting the replica positions),
with  $g_k$ not negligible for $k\le \xi_C^{-1},$  one has to impose $e^{i\,k\,a}\simeq 1,$
$a$ being the typical distance between the replicas.
In other terms, in (\ref{IntSpin}) the systems have to be coupled only with bath modes with
$k\ll a^{-1}.$
Now the whole Hamiltonian $H_{SB}=H_S+H_B+H_I,$ can be written by means of the global operators
$S^\alpha=\sum_{i=1}^N\sigma^\alpha_i\,(\alpha=\pm,z).$
In particular, the interaction reads
\begin{equation}
H_I=\sum_k(g_k\,S^+\,b_k+f_k\,S^-\,b_k^\dagger+h_k\,S^z\,b_k+\mbox{h.c.}).
\label{Spin}
\end{equation}
In such a case only the global generators  $S^\alpha$ are effectively coupled with the environment,
 whereby only collective
 coherent modes of
$\cal R$ are involved in the system dynamics.
Despite this simplification, the model described by $H_{SB}$
is in  general a  non-integrable interacting system, and therefore non trivial.
The exact eigenstates of $H_{SB}$ are generally given by highly entangled states of $\cal R$ and the bath.
Nevertheless since the $S^\alpha$'s span an algebra { isomorphic} with $sl(2),$
for $N$ even {\sl one can build a family of eigenstates of $H_{SB}$ given by simple tensor products}.
For $N=2$ let us consider the singlet state $|\psi\rangle=2^{-1/2}(|01\rangle-|10\rangle)$
(in a obvious binary notation): since $S^\alpha\,|\psi\rangle=0,\,(\alpha=\pm,z)$
one has that for {\sl every} $|\psi_B\rangle\in{\cal H}_B$ the  state $|\psi\rangle\otimes|\psi_B\rangle$
is annihilated by the interaction Hamiltonian. Moreover, it is a $H_{SB}$-eigenstate iff $|\psi_B\rangle$ is
a $H_B$-eigenstate, namely  
$|\psi_B\rangle$ has the form
$|\psi_B\rangle=\prod_{j} b_{k_j}^\dagger\,|0\rangle_B\equiv |K\rangle,$
where $K=(k_1,\ldots,k_n)$ denotes a $n$-tuple of wave vectors $k$ ($n\in {\bf N}$).
For $N>2$ (even) the existence of states $|\psi_j^{(N)}\rangle$
behaving like the singlet $|\psi\rangle$ is
ensured by the elementary $sl(2)$ representation theory.
The irreducible representations (irreps) ${\cal D}_j$ of $sl(2)$ are labelled by the total angular momentum
eigenvalue $j$ and are $2\,j+1$-dimensional.
When $j=0$ one has $1$-dimensional representations. The corresponding states (singlets)
are the many-qubit generalization of $|\psi\rangle.$
In general given a (reducible) representation $\cal D$ of ${sl}(2),$ one has
the Clebsch-Gordan decomposition in terms of the ${\cal D}_j$'s
\begin{equation}
{\cal D}^{\otimes\, N}=\bigoplus_{j\in{\cal J}} n_j {\cal D}_j,
\label{CG}
\end{equation}
the integer  $n_j$ being the multiplicity with which ${\cal D}_j$ occurs in the
resolution of ${\cal D}.$
The $S^\alpha$'s realize a (reducible) 
representation ${\cal D}_{1/2}^{\otimes\,N}$ of $sl(2)$ in ${\cal H}_S\cong ({\bf{C}}^2)^{\otimes\,N},$
that is  the $N$-fold tensor product of the (defining) $2$-dimensional representation 
${\cal D}_{1/2}.$
The Clebsch-Gordan series  reads for $N=2,\,4,\,6$
\begin{eqnarray}
{\cal D}_{1/2}^{\otimes\,2} &=& {\cal D}_1\oplus {\cal D}_0,\quad
{\cal D}_{1/2}^{\otimes\,4}= {\cal D}_2\oplus 3\,{\cal D}_1\oplus 2\,{\cal D}_0, \nonumber\\
{\cal D}_{1/2}^{\otimes\,6}&=&{\cal D}_3\oplus 5\,{\cal D}_2\oplus 9\, D_1\oplus 5\, {\cal D}_0.
\nonumber
\end{eqnarray}
Therefore, if $n(N)$ denotes the multiplicity
of the $j=0$ representation, one has 
 $n(2)=1,\,n(4)=2,\;n(6)= 5.$  
Let ${\cal C}_N$ be the $n(N)$-dimensional space spanned by the singlets: it is immediate 
-- by reasoning as in the $N=2$ case -- to verify 
that if $|\psi^{(N)}\rangle\in{\cal C}_N$ then $\forall |\psi_B\rangle\in{\cal H}_B$ one has 
$H_I\,|\psi^{(N)}\rangle\otimes|\psi_B\rangle=0.$
>From this property follows  the  result:\\
{\bf{Theorem 1 }}
 Let
${\cal M}_N$ be the manifold of  states  built over
the singlet space ${\cal C}_N.$
If  $\rho=\sum_{ij} R_{ij} |\psi_i^{(N)}\rangle\langle\psi_j^{(N)}|\in{\cal M}_N,$
then for any  initial bath state $\rho_B$ one has $L_t^{\rho_B}(\rho)=\rho,\; \forall t>0$.\\
{\it Proof}.\\
Let
${\rho_B}=\sum_{K^\prime, K} R_{K^\prime K} |\,K^\prime
\rangle\langle K|,$
and $\rho=\sum_{ij}\rho_{ij} |\psi^{(N)}_i\rangle\langle\psi^{(N)}_j|.$
Then, if $\rho(t)=U(t)\,\rho\otimes\rho_B\,U^\dagger(t),$
\begin{eqnarray}
\rho(t) &= & \sum_{ij,K^\prime K} \rho_{ij} R_{K^\prime K} \times \\ \nonumber & &U(t)\, |\psi^{(N)}_i\rangle\otimes
|K^{\prime}\rangle\,
(\langle\psi^{(N)}_j|\otimes\langle K\,|)U^\dagger(t)  \\ \nonumber
&=& \sum_{ij,K^\prime K} \rho_{ij} R_{K^\prime K} |\psi^{(N)}_i\rangle\otimes
|K^\prime\rangle\,\times \\ \nonumber
& & \langle\psi^{(N)}_j|\otimes\langle K\,|
e^{-i\,(E_{K^\prime} -E_K)\,t},
\end{eqnarray}
and taking the trace over the bath one gets
\begin{eqnarray}
\rho_t &=&\sum_{ij} \rho_{ij}|\psi^{(N)}_i\rangle\langle\psi^{(N)}_j|\times \\ \nonumber
& &\sum_{K^\prime K} R_{K^\prime K} e^{ -i\,(E_{K^\prime} -E_K)\,t}\mbox{tr}^B
 |\,K^\prime\rangle\langle K\,|\\ \nonumber
&=&\sum_{ij} \rho_{ij}|\psi^{(N)}_i\rangle\langle\psi^{(N)}_j|\sum_K R_{KK}=\rho,
\end{eqnarray}
where we used $\mbox{tr}^B(|\,K^\prime\rangle\langle K\,|)=\delta_{K^\prime K},$
and $\sum_K R_{KK}= \mbox{tr}^B\rho_B=1$ $\bullet$\\
The result stated by  Theorem 1 can be rephrased
in the following way which emphasizes its strength:
in the manifold of the states over ${\cal H}_S$ there exists
a submanifold ${\cal M}_N$ of fixed points (stationary states)
 of the  Liouvillian evolution.
The dynamics over ${\cal M}_N$ is therefore {\sl a fortiori} unitary.
Notice that this result relies only on algebra-theoretic properties and
not on any "perturbative" assumptions; in other words it holds for arbitrary
strength of the system-bath coupling.
This suggests the possibility of encoding in ${\cal M}_N$
 decoherence-free information,  namely the states of ${\cal M}_N$
realize a {\sl noiseless quantum code}.
For example  a (non-orthogonal) basis of ${\cal C}_4$ is
\begin{eqnarray}
|\psi_1^{(4)}\rangle &=& 2^{-1}(|1001\rangle-|0101\rangle+|0110\rangle-|1010\rangle), 
\nonumber\\
|\psi_2^{(4)}\rangle &=& 2^{-1}(|1001\rangle-|0011\rangle+|0110\rangle-|1100\rangle).
\nonumber
\end{eqnarray}
Orthonormalizing $|\psi_j^{(4)}\rangle,\,(j=1,2)$ one generates  a {\sl noiseless} qubit.
\\
It is remarkable that this result can be considerably generalized in many respects.
In the sequel we shall discuss such generalizations with no proofs;
the mathematical details will be given elsewhere \cite{ZARA}.
Basic ingredients are the concept of {\sl dynamical algebra} \cite{SOBI}
and   the standard Lie-algebra representation theory tools \cite{YOU}.
In what follows by dynamical algebra ${\cal A}_S$ of a system, with Hamiltonian $H\in\mbox{End}({\cal H}),$
we mean  the minimal Lie subalgebra of $gl({\cal H}),$ such that i) $H\in{\cal A}_S,$ ii) $H$ can be cast
in  diagonal form (i.e. linear combination of the
Cartan generators) by means of a  Lie algebra inner
 automorphism  $\Phi\colon {\cal A}_S\rightarrow {\cal A}_S$
(generalized Bogolubov rotation).\\
A system $S$ endowed with the dynamical algebra ${\cal A}_S$ 
with  Chevalley basis
$ \{e_\alpha,\, e_{-\alpha},\, h_\alpha\}_{\alpha=1}^r $,
can be thought of as
a collection of elementary excitations generated over the "vacuum"
by the raising  operators $e_\alpha$ of ${\cal A}_S.$
These excitations are  destroyed by the lowering generators
$e_{-\alpha}=e_\alpha^\dagger.$
The Cartan (abelian) subalgebra
 spanned by the
 $h_\alpha$'s
 acts diagonally. 
The $sl(2)$ (qubit) case corresponds to $r=1,$ the $e_\alpha$'s ($e^\dagger_\alpha$'s)
are the analog of $\sigma^-$ ($\sigma^+$) whereas the  $h_\alpha$'s correspond to $\sigma^z.$
The Hamiltonian can be written, in view of   ii) above, in diagonal form as
   $H=\sum_{\alpha=1}^r\epsilon_\alpha h_\alpha.$
We consider now $N$ non-interacting replicas of $ S$.
The Hilbert space  becomes 
$
{\cal H}_S={\cal H}^{\otimes\,N},$ with $\mbox{dim}({\cal H}_S)=d^N.$
As in the qubit case it  is useful to introduce   the global operators 
$X_\alpha\equiv
\sum_{j=1}^{N}  x_\alpha^j,$ where $x_\alpha^i$ acts as $x_\alpha\in{\cal A}_S$
in the $i$-th factor of the tensor product, and as the identity in the remaining factors. 
The operators 
$\{\,E_\alpha,\,E_{-\alpha},\,H_{\alpha}\}$ span an algebra isomorphic with ${\cal A}_S.$
The global Hamiltonian of the register can be written then in terms of the generators $H_\alpha$   of the Cartan
subalgebra of ${\cal A}_S$ as
$
H_S=\sum_{\alpha=1}^r \epsilon_\alpha\,H_\alpha.
$
We assume that the system-bath interaction couples directly the bosonic modes
with the elementary excitations of the system. 
The interaction   Hamiltonian has  the
form, analog to that of  equation (\ref{Spin}), 
$$
H_I=
\sum_{k\alpha}\tau_{\alpha}\,(g_{k}^\alpha \,E_\alpha\,b_k+
 f_{k}^\alpha\,E^\dagger_\alpha\,b_k^\dagger+ h_{k}^\alpha H_\alpha\,b_k\,+\mbox{h.c.}),
$$
where we have already assumed the replica-symmetry of the coupling functions.
The elementary processes associated with this $H_I$
have the same interpretation as in the  qubit case.
As far as our basic result is concerned the assumption -- physically motivated --
that $S$ is bilinearly
coupled with the bath by the Chevalley basis operators of the ${\cal A}_S^i$'s
is not restrictive.
Indeed, if one were given as initial data not the dynamical algebra ${\cal A}_S,$
but the system operators coupled with the environment as well as $H$
one could reconstruct ${\cal A}_S$ by closing all possible commutation relations.
In the generic case the algebra ${\cal A}_S$ thus generated 
turns out to be semisimple and acts irreducibly on $\cal H.$
Since the global  operators span an
algebra {\sl isomorphic} with  ${\cal A}_S,$ one  can use the  ${\cal A}_S$
 representation theory  to split ${{\cal H}_{SB}}= {\cal H}_S\otimes {\cal H}_B$
 according to the irreps of ${\cal A}_S.$
In the following without loss of generality, we let  
${\cal A}_S\equiv sl(r+1),$ and let ${\cal D}$ denote
the defining representation of ${\cal A}_S$ in $\cal H$
($d=\mbox{dim}\, {\cal H}= r+1$).
We need to consider 
the Clebsch-Gordan series of the $N$-fold tensor product representation
of ${\cal A}_S$ in ${\cal H}^{\otimes\,N}.$ 
It has the same form of (\ref{CG}),
the set ${\cal J}$ being  now the  label set for the irreps of $sl(r+1),$
and $n_j$ the multiplicity of the irrep ${\cal D}_j.$
An easy way to compute the GC series is to resort to the Young diagrams
which relate the representation theory of $sl(r+1)$ with that of the symmetric group
${\cal S}_N$ \cite{YOU}.
Each Young diagram $\cal Y$ is associated with an irrep of ${\cal S}_N.$ 
Indeed,
if $|\psi\rangle=\otimes_{j=1}^{N}|\psi_j\rangle$ is a basis vector of ${\cal H}^{\otimes\,N},$
 the formula 
 $\sigma\,|\psi\rangle=\otimes_{j=1}^{N}|\psi_{\sigma(j)}\rangle$
defines, for any $\sigma\in{\cal S}_N,$
 by linear extension, a natural  ${\cal S}_N$-action over ${\cal H}^{\otimes\,N}.$
The multiplicities $n_j$ are the dimensions of the ${\cal S}_N$-irreps associated
with  ${\cal Y}.$
  The dimension $d_j$ of ${\cal D}_j$ is given by the number of different
Young tableaux that one can obtain from ${\cal Y},$ and is equal to the multiplicity
 of the associated ${\cal S}_N$-irrep.
For $N=r+1$ one finds, with multiplicity one, 
 the (fundamental) 
antisymmetric representation ${\cal D}_A,$ associated with the $(r+1,\,1)$ Young diagram
 with just one column of  $N$ boxes 
(we use the notation $(n,m)$ for the {\sl rectangular } Young diagram
with $n$ rows and $m$ columns).
${\cal D}_A$ is one-dimensional, and given by the vector
$$
|\psi_A\rangle={N!}^{-1/2}\,
\sum_{\sigma\in{\cal S}_N}(-1)^{|\sigma|} \sigma \otimes_{j=1}^N|j\rangle,
$$
 $\{|\,i\rangle\}_{i=1}^N$ being  a basis for ${\cal H},$ and $|\sigma|$ denoting the parity of $\sigma.$
Now we observe that, since  $|\psi_A\rangle$ is a $sl(r+1)$-{\sl singlet},
 one must have $H_\alpha\,|\psi_A\rangle= E_\alpha\,|\psi\rangle=
E_{-\alpha}\,|\psi\rangle=0,
(\alpha=1,\ldots,r).$
Therefore for $|\psi_B\rangle$   { any} vector of ${\cal H}_B,$  $|\psi_A\rangle\otimes |\psi_B\rangle$ 
is {annihilated by the interaction Hamiltonian} and is an eigenstate of $H_S+H_B$ iff $|\psi_B\rangle$ 
is an eigenstate
of $H_B.$
More generally for $N=m\,(r+1),\,(m\in{\bf N})$ one has the $(r+1,\,m)$ Young diagram 
with multiplicity $n(N)$,  still corresponding
to  one-dimensional representations of $sl(r+1)$.
Let $|\psi^{(N)}_j\rangle,\,(j=1,\ldots,n(N))$ denote 
the associated vectors, then, reasoning as above, we have
that $|\psi^{(N)}_j\rangle\otimes \,|K\rangle_B$ is an eigenstate of $H_{SB}$ with
eigenvalue $E_K=\sum_{j} \omega_{k_j}.$
With the procedure described  above we have therefore built an infinite family
of {\sl exact} eigenstates of the interacting Hamiltonian $H_{SB}$ that are given by simple 
tensor products. This allows us to  state straightforwardly the following generalization of 
Theorem 1:\\
{\bf Theorem 2}
Let ${\cal C}_N=\mbox{span}\{ |\psi^{(N)}_j\rangle\,|\,j=1,\ldots,n(N)\},$ with $N=0\, 
\mbox{mod} (r+1),$
and ${\cal M}_N$ the manifold of the  states over ${\cal C}_N.$ Then: 
if $\rho\in {\cal M}_N,$  for any state $\rho_B$ over ${\cal H}_B$ one has
$L^{\rho_B}_t\,\rho =\rho.$\\
The proof proceeds as in the qubit case.
The code is nothing but ${\cal C}_N$ itself.
For $N=2\,(r+1)$ one has $n(N)=2$ and a single {\sl qubit} can be encoded.
As far the encoding efficiency
is concerned, we observe that in the $r=1$ case one has
$n(N)=N![(N/2)!(N/2+1)!]^{-1},$ ($N$ even) from which follows, for large N, the asymptotic form
$\log_2 n(N)\simeq N -3/2\,\log_2 N.$
The latter equation tells us that, for large replica number, one has an enconding efficiency
$N^{-1}\,\log_2 n(N)$  approximately of one qubit per replica, whereas the fraction
$2^{-N}\,n(N)$ of the Hilbert space occupied by the code is vanishingly small.
In the general case $r>1$ the multiplicities  $n(N)$ are the  Littlewood-Richardson
coefficients \cite{LIRI}.
A few important remarks extending  Theorem 2 follow.
i)
When only  the dephasing terms
 are present, 
due to the fact that the resulting model can be diagonalized
by a  unitary transformation in each
${\cal A}_S$-weight space \cite{WEI},
if $\rho$ is a state over ${\cal H}_S(\lambda)$ then $L^{\rho_B}_t\rho =\rho.$
This latter result, in its simplest form (i.e. $r=1$) can be found in \cite{PALMA}
and \cite{LUGU}.
Notice that this model does not take into account the amplitude errors
induced by the bath.
ii) We can allow also for interactions $H_{SS}$ between replicas, provided they leave
 ${\cal C}_N$ invariant. For example it would be sufficient  that  ${\cal A}_S$
were a symmetry algebra for $H_{SS}.$
There results $L^{\rho_B}_t\rho= U_S(t)\,\rho\,U_S^\dagger(t),$
where $U_S(t)=e^{-i\,H_{SS}\,t},$ therefore the Liouvillian dynamics is still
unitary but no longer trivial.
iii) Since ${\cal C}_N$ is an irreducible ${\cal S}_N$ representation space,
 the theorem still holds (with non-trivial unitary evolution)
if the Hamiltonian $H_S$ and the system operators coupled with the bath
belong to the symmetric subspace of $\mbox{End}({\cal H}_S).$
>From the physical point of view this means that we can  allow
for replica-replica and replica-bath interactions 
involving many excitations (powers of the $e^i_\alpha$'s)
provided all the replicas are treated symmetrically.\\
We expect that
if the  key assumption of a replica-symmetric coupling with the bath is slightly
violated -- for example the system is coupled with modes with wavelengths shorter than the inter-replica distance  -- 
 the proposed encodings  have a low error  rate, in analogy with  the "sub-decoherent" states
in \cite{PALMA}.\\
In summary, we have shown that for open quantum systems, made of $N$  replicas
of a given system $S,$ coupled with a common environment in a replica-symmetric fashion, one can build
-- for sufficiently large $N$ -- a 
subspace ${\cal C}_N$  of ${\cal H}^{\otimes\,N}$ that does not get entangled with the environment.
The whole  class of (possibly non-linear) replica-replica interactions
which leave ${\cal C}_N$ invariant togheter with the replica-symmetric
system-bath interactions (which  possibly
annihilate ${\cal C}_N$) is consistent with this scheme.
Such subspace is  nothing but the singlet sector 
of the dynamical algebra ${\cal A}_S$ of $S,$ direct sum of the
one-dimensional representations  of  ${\cal A}_S$.
This elegant result allows us, in  principle, to design noiseless (i.e.
dissipation/decoherence free)  quantum codes.
>From the point of view of the practical implementation the difficulties
one may expect to face with  these codes 
depend on the limitations 
inherent with  the code-words preparation and on  the large bath coherence length
required.
The question of the code stability, in the case in which the latter requirement is not satisfied,
can be addressed in the framework of the Liouville-von Neumann equation formalism \cite{ZA}.
Another  open question is whether the approach discussed may be possibly extended
to the case when $\cal H$ is infinite dimensional. 
Work is in progress along these lines.\\ 
Discussions with R. Zecchina are  acknowledged.
P.Z. thanks C. Calandra
for  hospitality at the  University of Modena
 and  Elsag-Bailey
for financial support.

\end{multicols}

\begin{references}
\bibitem{WEI}
The Hilbert subspace ${\cal H}_S(\lambda)$ of ${\cal H}_S$
 is called  ${\cal A}_S$-weight space
if $|\psi\rangle\in{\cal H}_S(\lambda)\Rightarrow H_\alpha\,
|\psi\rangle= \lambda(H_\alpha)\,|\psi\rangle$.
\end{references}
\end{document}